# Normative brain mapping of interictal intracranial EEG to localise epileptogenic tissue

Peter N Taylor[1,2,*,†], Christoforos A Papasavvas[1], Thomas W Owen[1], Gabrielle M Schroeder[1], Frances E Hutchings[1], Fahmida A Chowdhury[2], Beate Diehl[2], John S Duncan[2], Andrew W McEvoy[2], Anna Miserocchi[2], Jane de Tisi[2], Sjoerd B Vos[2], Matthew C Walker[2], Yujiang Wang[1,2,*,†]

[†]These authors contributed equally to this work.

# Abstract

The identification of abnormal electrographic activity is important in a wide range of neurological disorders, including epilepsy for localising epileptogenic tissue. However, this identification may be challenging during non-seizure (interictal) periods, especially if abnormalities are subtle compared to the repertoire of possible healthy brain dynamics. Here, we investigate if such interictal abnormalities become more salient by quantitatively accounting for the range of healthy brain dynamics in a location-specific manner.

To this end, we constructed a normative map of brain dynamics, in terms of relative band power, from interictal intracranial recordings from 234 subjects (21,598 electrode contacts). We then compared interictal recordings from 62 patients with epilepsy to the normative map to identify abnormal regions. We hypothesised that if the most abnormal regions were spared by surgery, then patients would be more likely to experience continued seizures post-operatively.

We first confirmed that the spatial variations of band power in the normative map across brain regions were consistent with healthy variations reported in the literature. Second, when accounting for the normative variations, regions which were spared by surgery were more abnormal than those resected only in patients with persistent post-operative seizures ($t=-3.6$, $p=0.0003$), confirming our hypothesis. Third, we found that this effect discriminated patient outcomes ($AUC=0.75$ $p=0.0003$).

Normative mapping is a well-established practice in neuroscientific research. Our study suggests that this approach is feasible to detect interictal abnormalities in intracranial EEG, and of potential clinical value to identify pathological tissue in epilepsy. Finally, we make our normative intracranial map publicly available to facilitate future investigations in epilepsy and beyond.

**Author affiliations:**

1. CNNP Lab (www.cnnp-lab.com), Interdisciplinary Computing and Complex BioSystems Group, School of Computing, Newcastle Helix, Newcastle University, NE4 5TG, UK

2. UCL Queen Square Institute of Neurology & National Hospital for Neurology and Neurosurgery (NHNN), Queen Square, London WC1N 3BG, UK

*Correspondence to: Peter Taylor & Yujiang Wang

Full address: Urban Science Building, Newcastle Helix, Newcastle upon Tyne, UK

Email: peter.taylor@newcastle.ac.uk yujiang.wang@newcastle.ac.uk

# Introduction

Abnormal electrographic activity is a hallmark of many neurological disorders. In focal epilepsy, ictal (seizure) periods commonly display clear pathological dynamics, which is clinically used to localize epileptogenic tissue. However, studies have suggested that interictal dynamics may also hold useful complementary information to identify epileptogenic tissue. For example, interictal spikes, sharp waves, and high frequency oscillations have all been suggested as putative markers.[1–8]

Grossly abnormal interictal events, such as interictal spikes, can often be identified visually or algorithmically. However, existing techniques may struggle to distinguish more subtle aberrations from the vast repertoire of possible healthy brain dynamics. Example healthy brain dynamics include beta oscillations, commonly seen in motor areas,[9,10] and gamma activity in occipital and temporal areas.[11,12] Other spatial profiles include alpha oscillations in occipital and parietal areas,[11,13] delta in the temporal lobe,[12–14] and theta in superior frontal areas.[15,16] In this work, we suggest that neural activity in these frequencies may also represent pathological activity if it occurs in brain regions that do not normally feature these frequencies. Conversely, a lack of power in typical frequencies of a particular brain region may also indicate pathological activity. Thus, identifying such subtle pathological activities requires the consideration of the spatial distribution of 'normal' electrographic activity.

One approach to account for normal spatial variations is to construct a normative map, which describes the healthy spatial profile and ranges of the feature of interest (in this case the band power of different frequency bands). Such an approach is common and well-accepted in neuroimaging of brain structural abnormalities: patients are often normalised against healthy controls to highlight abnormal brain morphology[17] or connectivity.[18,19] However, for invasive recordings using intracranial EEG, data from healthy controls are not available. Instead, recent studies suggested using intracranial EEG recorded from areas outside of the putative seizure-generating tissue in patients with epilepsy.[13,20] Specifically,[13] conclude that this approach yields a normative map of brain dynamics that is consistent with data from animal models and other recording modalities.[21] also motivated the use of normative maps using intracranial EEG.

In this study, we therefore follow this proposed approach to generate a normative map of band power across the brain using intracranial recordings from 234 subjects with 21,598 recording contacts from outside the seizure onset and initial propagation zone. We first quantify the spatial distributions of normative band power and confirm agreement with previous data. Then, using a separate cohort of 62 patients with epilepsy, we show that accounting for the normative map allows us to identify epileptogenic tissue and subsequently predict patient surgical outcomes.

# Methods

## Patients

Two main cohorts were studied here. The RAM normative cohort consisted of 234 subjects with epilepsy undergoing presurgical evaluation with intracranial EEG to localise seizure onset. As part of the intracranial EEG monitoring, the subjects were also participating in an experimental study on memory (data collected up to year 3; http://memory.psych.upenn.edu/RAM). As stated in the project's website "Informed consent has been obtained from each subject to share their data, and personally identifiable information has been removed to protect subject confidentiality". The original research protocol for data acquisition was approved by the relevant bodies at the participating institutions. Furthermore, the University Ethics Committee at Newcastle University approved the analysis of this dataset (Ref: 12721/2018). The normative recordings were obtained in the preparatory phase, several minutes before a memory task.

The UCLH epilepsy cohort consisted of 62 patients with epilepsy undergoing presurgical evaluation with invasive intracranial EEG to localise seizure onset. All patients had pre-surgical, pre-implantation T1-weighted (T1w) MRI. All patients had either CT or T1w MRI whilst implanted electrodes were in place. The majority of patients had post operative T1w MRI (N=61). For the single patient without post-operative MRI, the detailed surgery report described the brain areas resected. At follow-up of 12 months, 33 patients were free of disabling seizures and 29 had persistent seizures. Follow-up outcomes were defined as described previously according to the ILAE classification.[22] A subset of this cohort has been studied previously.[23] All data were anonymised and exported, then analysed under the approval of the Newcastle University Ethics Committee (2225/2017). Detailed patient metadata are shown in Supplementary Information S1 and summarised in table 1. No significant differences were present between outcome groups in age, sex, lobe of resection, side of resection, or number of electrode contacts.

***Table 1: Summary of patient data.***

| | ILAE1,2 | ILAE3+ | Test statistic |
|---|---|---|---|
| N (%) | 33 (53%) | 29 (47%) | |
| Age (mean,SD) | 32.3 (10.7) | 33.0 (8.8) | p=0.8017, t=-0.2522 |
| Sex (M,F) | 15,18 | 17,12 | p=0.3, $\chi^2$=1.07 |
| Temporal, extratemporal | 21,12 | 15,14 | p=0.34, $\chi^2$=0.8995 |
| Side (Left, Right) | 18,15 | 16,13 | p=0.96, $\chi^2$=0.00024 |
| Number of electrode contacts (mean, SD) | 71.1 (24.3) | 65.9 (23.3) | p=0.3984, t=0.8505 |

## MRI processing for electrode localisation and resection delineation

Electrode contacts for all subjects were localised to regions of interest defined according to a parcellation. To ensure robustness of our findings we investigated four separate parcellations at different resolutions where higher resolutions are subdivisions of lower resolutions. These parcellations have been described previously[24] and have been used for normative intracranial analysis.[20] Due to different levels of available data, our technique for localisation of electrode contacts to regions differed slightly between the RAM and UCLH datasets. In the RAM data, electrode contact locations are publicly available as Talairach space coordinates, which we converted to MNI space.[25] We next reconstructed an MNI space brain using FreeSurfer, matched each of the four parcellations to that surface using mri_surf2surf, obtained the labels and matched each contact to the closest volumetric region of interest (minimum euclidean distance using custom code in matlab). For UCLH data we performed broadly the same procedure but performed the processing in native space. Performing native space processing was possible as the pre-operative T1w MRI was available along with the CT/MRI scan to mark electrode contacts as described previously.[23,26] To identify which regions were removed/spared by surgery we linearly registered the post-operative T1w scan to the pre-operative scan and manually delineated the resected tissue as a mask described previously.[23,27] Electrode contacts were defined as removed if they were within 5mm of the mask as in our prior work.[23] In each patient, regions were defined as removed if >25% of contacts within the region were removed, otherwise regions were considered spared.

## Intracranial EEG data and processing

To create a normative baseline of intracranial EEG (iEEG) spectral properties we used the RAM dataset, and extracted 70 seconds of iEEG recording from relaxed wakefulness (shortly before a memory task) for each subject. We excluded channels that were labelled as seizure onset zone, early propagation zone, brain lesions, or bad contacts. The extracted EEG signals from the remaining channels were visually inspected for recording artefacts, and recording channels located in white matter were also excluded, resulting in a final set of 21,598 channels across 234 subjects.

We further used a separate iEEG dataset from UCLH to compare and score against the normative baseline. Again, we retrospectively extracted 70 seconds of interictal iEEG recording for each subject, at least 2h away from seizures. Where possible, the recording was obtained at around 2pm in the afternoon to maximise the likelihood of wakefulness. Due to the retrospective design, it was not possible to determine the exact brain state. To demonstrate robustness, we also present results for two further time segments at least 2h away from seizures and 4h away from other time segments at around 9am and 7pm where possible. For the UCLH dataset, we included all grey matter channels (i.e. even those in seizure onset zone, propagation zone, and irratative zones). We only excluded artefactual channels and recording channels in white matter, resulting in 4256 channels across 62 patients.

All EEGs were downsampled to 200 Hz for the RAM dataset before creating the normative map. In the UCLH data, we had a mixture of sampling frequencies (2 subjects at 256Hz, 52 subjects at 512 Hz, 8 subjects at 1024 Hz, and 1 subject at 2048 Hz). After applying a common average reference to all recordings in all subjects, we estimated the power spectral density with Welch's method (2 s window, 1 s overlap, Hamming window) in each 70 s recording. The average band power within five frequency bands of interest were then calculated using the 'bandpower' function in matlab. The following ranges were defined, delta ($\delta$ 1-4 Hz), theta ($\theta$ 4-8 Hz), alpha ($\alpha$ 8-13 Hz), beta ($\beta$ 13-30 Hz) and gamma ($\gamma$ 30-80 Hz). In the gamma band, data between 47.5 Hz to 52.5 Hz and 57.5 Hz to 62.5 Hz was excluded to avoid power line artifacts in both the US and UK recordings. Band power estimates were then $\log_{10}$ transformed and normalised to sum to 1 for each contact (i.e. L1 norm). These transformed and normalised values represent the relative band power used throughout results. Each subject therefore has a value of relative band power assigned to each contact and each frequency band.

For the UCLH dataset, the clinical team additionally provided information on if any channels displayed inter-ictal spikes (at any point during the recording). This information was used later as a baseline measure, and to demonstrate robustness of our results.

## Normative map generation

To obtain a normative distribution of relative band power in a particular frequency band and brain region, we first assigned each electrode contact from each subject in the RAM dataset to a grey matter region, as described above. One contact can only be assigned to a single (nearest) region. If multiple contacts from the same subject were assigned to the same region, then we averaged the relative band powers to obtain single values of relative band power per region and frequency band per patient. If zero contacts were assigned to a region in a particular subject, then the region was considered to have no coverage and the relative band powers were set to NaN (not a number) for that subject and region. The normative distribution of relative band power in a region (in a particular frequency band) was then obtained as the distribution of relative band powers of all RAM subjects with coverage in that region. Coverage obtained in the normative map can be found in supplementary analysis 6.

To visualise the normative map, we plot the mean of the distribution of relative band powers in a particular region and frequency band across normative subjects (see Fig. 1).

## Scoring patients to the normative map

To score the UCLH patient cohort against the normative map, we followed a similar approach in mapping the electrodes to brain regions. Electrode contacts for a given patient were localised to a single brain region ($i$). Where multiple contacts localised to the same region the mean band power value across contacts was used. This allows estimation of the band power in a given region ($i$), in a given frequency band ($j$), for a given patient. To estimate the abnormality of a region's relative band power in the UCLH dataset from the normative map we computed the absolute z-score (eq. 1):

$$\left|z_{i,j}\right| = \left|\frac{x_{i,j} - \mu_{i,j}}{\sigma_{i,j}}\right| \quad (1)$$

where $i$ represents the region, and $j$ the frequency band of interest, $x$ is the band power value for an individual patient, $\mu$ and $\sigma$ are the mean and standard deviations of the band powers in the normative map.

In comparing the values between resected and spared regions for any patient in the UCLH dataset and frequency band, we used the distiguishability statistic ($D_{RS}$), which is the area under the receiver operating curve, and equivalent to the normalised Mann-Whitney U statistic (see).[23,28,29] A $D_{RS}$ value greater than 0.5 indicates that spared regions were more abnormal (higher absolute z-score) than resected regions, whereas $D_{RS}$ values below 0.5 indicates the opposite - i.e. resected regions were more abnormal.

## Statistical analysis

Hypothesising that resected regions would be more abnormal than spared regions in good outcome patients, we tested for $D_{RS} < 0.5$ in good outcome patients using a left tailed one sample t-test. In contrast, we hypothesised the opposite effect in poor outcome patients and tested $D_{RS} > 0.5$ using a one sample right tailed t-test. Finally, we hypothesised greater $D_{RS}$ values in poor outcome patients than good outcome patients, and tested with a two sample left tailed t-test.

Statistical significance is reported for $p < 0.05$ for reference. Effect sizes are reported throughout as t-statistics or as area under the receiver operating characteristic curve (AUC).

## Data and code availability

Preprocessed data and analysis code will be made available upon acceptance of the manuscript.

# Results

## Normative maps show spatial organisation of band power

We constructed normative maps of relative power in five frequency bands ($\delta$: 1-4Hz, $\theta$: 4-8Hz, $\alpha$: 8-13Hz, $\beta$: 13-30Hz, and $\gamma$: 30-80Hz). To construct the normative maps, we used 70 seconds of interictal intracranial EEG recordings from 21,598 electrode contacts outside of the seizure onset and initial propagation zone across 234 subjects. The 70 second segments were recorded while the subjects were awake and preparing for a cognitive task experiment. We derived the relative band power for five main frequency bands in all contacts. Each contact was then assigned to one of 128 regions of interest (ROIs) from the Lausanne scale60 atlas,[24] yielding a normative distribution of relative band power in each ROI.

The resulting normative maps of the mean relative band power for each frequency band are shown in Figure 1A. Several distinct patterns can be observed; for example, relative delta power is most prominent in the anterior temporal and anterior frontal regions, while relative alpha power is prominent in parietal and occipital regions. Note that lower frequencies generally have higher relative power (the colour axes scale differs for each frequency band in Fig. 1A). Finally, the overall gradient of the normative maps also display a striking symmetry between the left and right hemispheres. These normative spatial profiles are further quantified in supplementary analysis 5.

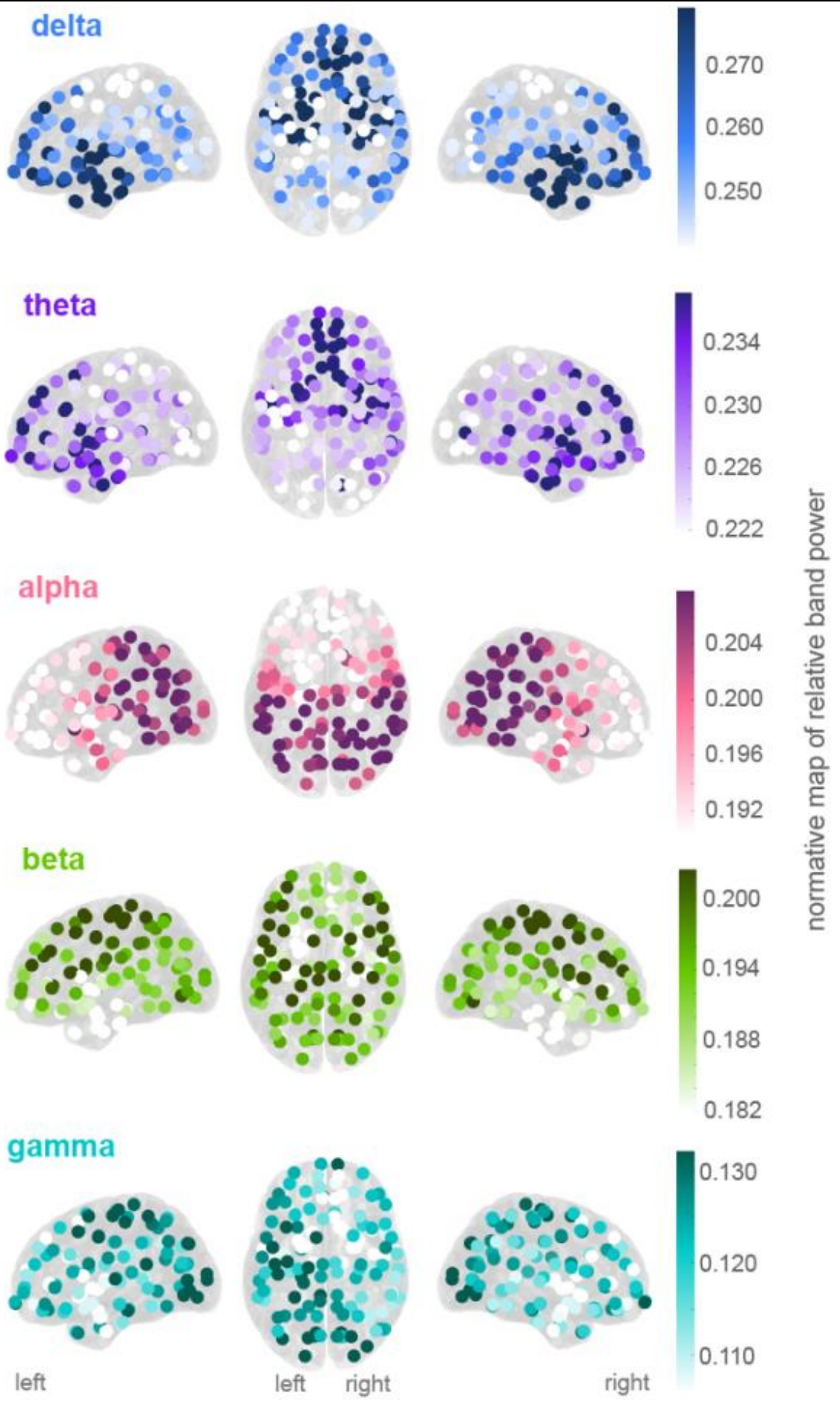

**Figure 1: Normative band power varies across regions.** Mean relative band power in each region for each of the five frequency bands of interest. The colour axes scale differs for each frequency band with generally higher power in lower frequencies.

## Normative maps highlight abnormalities in individual patients

We then turned our attention to a cohort of patients from UCLH with refractory focal epilepsy who underwent presurgical evaluation with intracranial EEG. We used the normative maps as a baseline to identify aberrations in each ROI for individual patients.

We use an example patient to illustrate the process. Patient ID 1216 had electrode contacts placed in the temporal, parietal and occipital lobes. Those electrodes were localised to corresponding ROIs (black circles in Figure 2A). We also show the interictal EEG time series of two example contacts in two different regions in Fig. 2A. The first region is the left middle temporal gyrus 2 (LMTG2), which is far away from the seizure onset zone in this patient. The second region is the left lateral occipital gyrus 2 (LLOG2), which is the seizure onset zone as determined by the presurgical evaluation.

On visual inspection, the two time series are not qualitatively different. However, following extraction of relative band power from the 70 s interictal recording for each of the two regions, and subsequent standardisation to the normative distributions in each frequency band (violin plots in Fig. 2B), the LLOG2 region showed substantial deviations, particularly in the theta band (absolute z-score of 2.99). In contrast, the LMTG2 region did not display any strong deviations in any frequency band ($\lesssim 1$ standard deviation away from the normative distributions).

We repeated the procedure of z-scoring all frequency bands in all ROIs relative to the corresponding normative distributions for example patient 1216. It is conceivable that different frequency bands are abnormal in different regions and subjects (Supplementary analysis 8). Therefore, to summarise these z-scores across frequency bands, we used the maximum absolute z-score as a measure of the regional level of aberration (Fig. 2C). Taking the maximum essentially summarised the level of interictal band power abnormality whilst allowing for region and subject-specific differences in terms of the frequency band. In this example patient, it is visually clear from Fig. 2C that the level of abnormality is highest in the LLOG2 region, but other occipital regions also presented with a high level of abnormality.

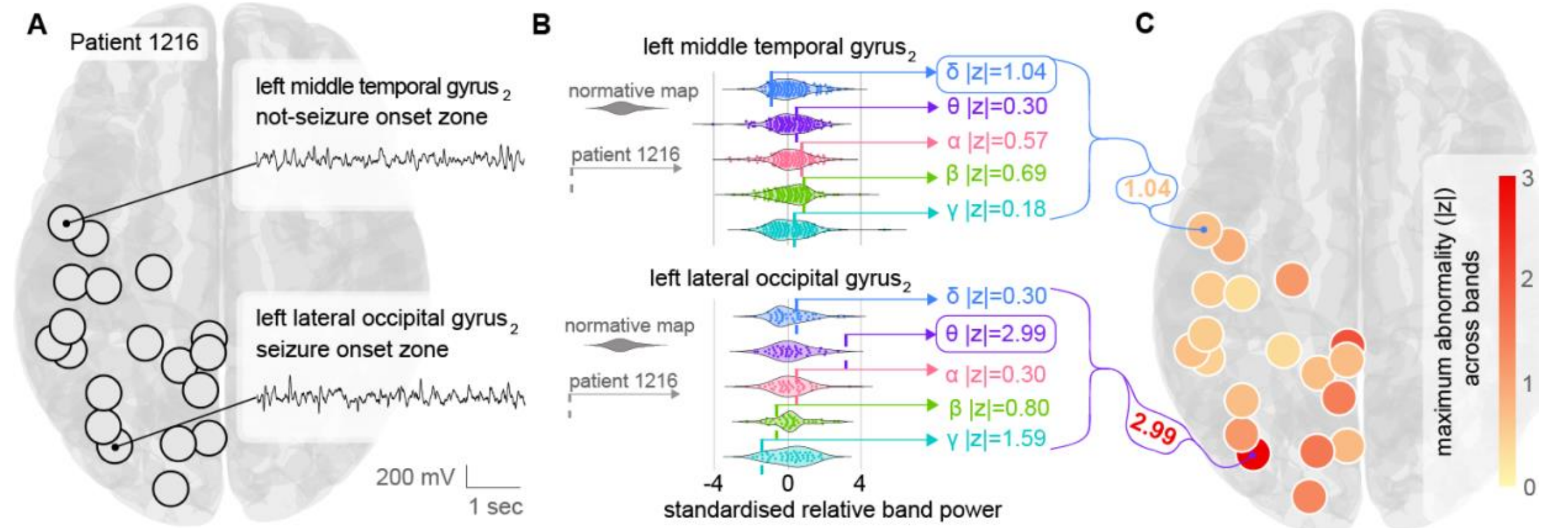

**Figure 2: Normative band power as a reference to detect abnormalities in individual patients. A.** Visualisation of the regions covered by the implanted electrodes in an example patient with epilepsy. 18 of the 128 regions were sampled by the electrode contacts in this patient (black circles). Time series from two example regions are shown that are without obvious epileptiform activity (inset). One example region (left lateral occipital gyrus 2) was the seizure onset zone in this patient. **B.** Relative band power for each of the two regions, across each frequency band is plotted for the normative data (coloured violin plot; each point is a normative subject). Data is standardised (mean subtracted and divided by standard deviation). Relative band power z-score for the patient 1216 is plotted as a vertical dashed line on the same scale. The z-scores indicates that the left middle temporal gyrus is normal in all frequency bands (maximum absolute z-score = 1.04). The left lateral occipital gyrus is more abnormal in theta (maximum absolute z-score = 2.99) and gamma (absolute z-score = 1.59). **C.** Maximum absolute z-score for each region plotted for the patient. Larger values indicate greater abnormality in any frequency band.

## Interictal band power abnormality distinguishes epileptogenic tissue

We next postulated that our measure of interictal band power abnormality of a region may serve as a marker of the region's epileptogenicity. We thus hypothesised that the surgical removal of regions with the greatest abnormalities would be associated with post-operative seizure freedom. In contrast, if abnormal regions remain after surgery, we expect to see persistent seizures after surgery. To address this hypothesis, we retrospectively identified which regions were resected by surgery and compared the level of abnormality between surgically resected and spared regions.

The example patient 1216 in Figure 3A,B is the same patient shown in Figure 2. The LLOG2 region was resected, along with other occipital regions. It is visually apparent that the resected regions (circled in black in figure 3B) appear substantially more abnormal than regions which were spared by surgery in this first example patient. The lower panel of Fig. 3B quantifies the difference between the resected and spared regions using the $D_{RS}$ metric that quantifies the Distinguishability of the Resected and Spared regions in an individual patient.[23,28,29] $D_{RS}$ values close to 0 indicate that resected regions are more abnormal than spared regions in that individual patient. In contrast, if $D_{RS}$ is close to 1, then spared regions are more abnormal than resected regions. A $D_{RS} = 0.5$ indicates that the resected and spared regions are indistinguishable in terms of the level of interictal band power abnormality. Example patient 1216 has a $D_{RS} = 0.14$ (Fig. 3B), indicating that regions removed by surgery were typically more abnormal than regions spared by surgery. This patient was subsequently seizure free upon follow-up.

Interictal band power abnormalities of a second example patient (ID: 910), derived using the same processing and normative analysis, are presented in Figure 3C,D. This patient had an anterior frontal lobe resection. Their resection involved the removal of areas with normal interictal band powers ($|z| \lesssim 1$), whilst highly abnormal regions remained in more posterior parts of the frontal lobe. Analysis using $D_{RS}$ confirms this finding with $D_{RS} = 0.96$, indicating that almost all spared regions were more abnormal than those resected. This example patient continued to have persistent post-operative seizures.

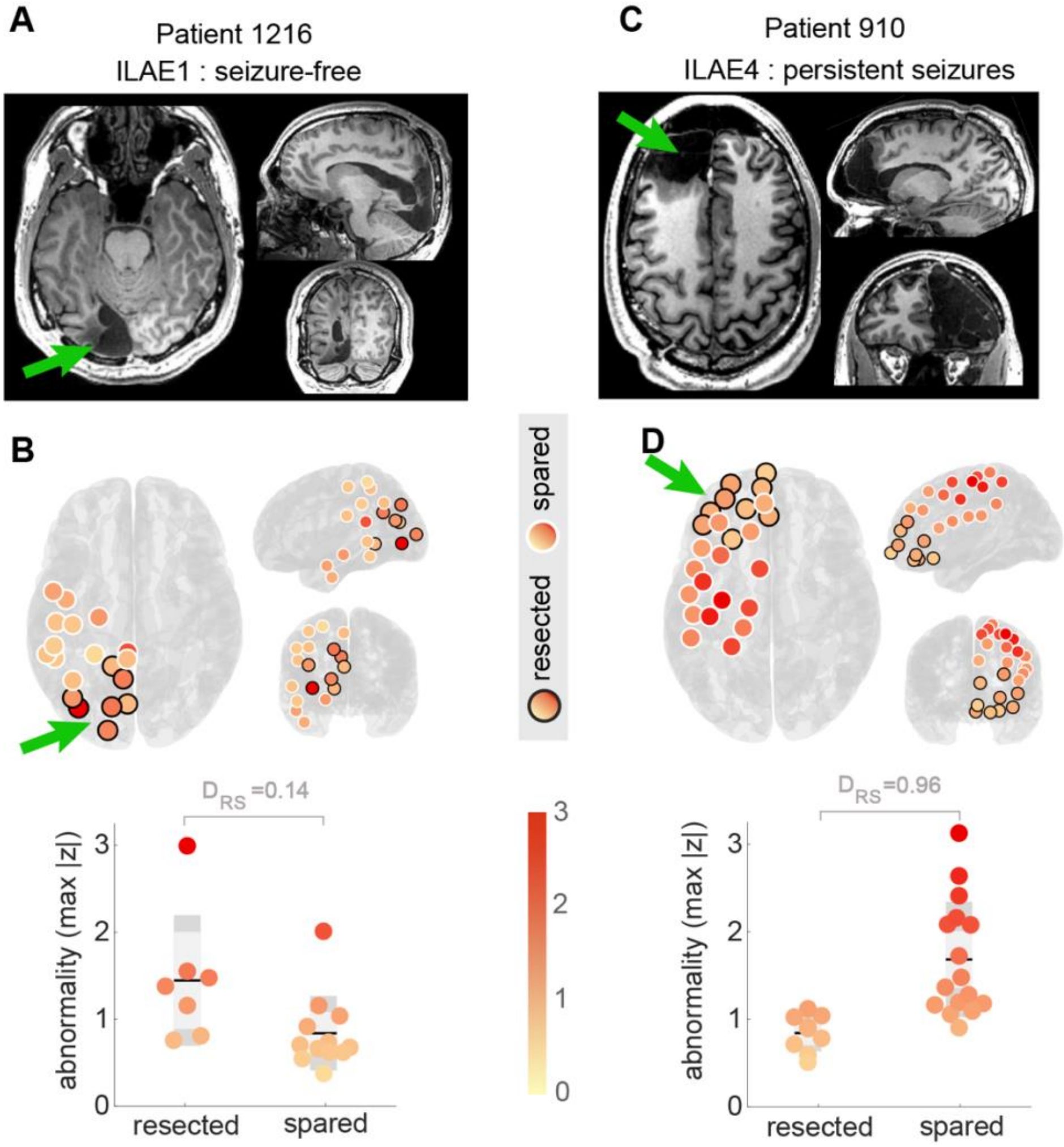

**Figure 3: Interictal band power abnormality as a marker of epileptogenic tissue in two example individual patients. A, C** Post-operative T1-weighted MRI scans showing the location of the resection as indicated by the green arrow. **B.** Replication of the patient in figure 2 with the regions that were later surgically resected circled in black. Non-resected regions are circled in white. A direct comparison and quantification in the lower panel shows resected regions to be more abnormal than spared. Each data point is a separate region. This patient was seizure free after surgery (ILAE1). **D.** Visualisation of data from a second patient with a frontal lobe implantation. Multiple abnormal regions were present outside the resection and spared by surgery. This patient had had continued post-operative seizures (ILAE4). In both patients the $D_{RS}$ metric quantified the difference between resected and spared regions in terms of their abnormality.

The two patients presented in Figure 3 suggest that the interictal band power abnormality measure may serve as a marker of epileptogenicity, and its ability to distinguish resected from spared tissue ($D_{RS}$) may subsequently be used to predict seizure-freedom after surgery. In Figure 4A,B we generalise those findings across a cohort of 62 patients, with each datapoint representing an individual patient. At a group level, patients with persistent seizures (ILAE3+) had substantially and significantly greater $D_{RS}$ values than those who were free of disabling seizures (ILAE1,2) (right tailed t-test p=0.0003, t=-3.6, AUC=0.75, see Fig. 4A,B). Furthermore, $D_{RS}$ values of patients with persistent seizures were substantially and significantly greater than 0.5, suggesting that abnormal regions were spared by surgery in ILAE3+ patients (p=0.0003, t=3.8, right tail t-test). Values of $D_{RS}$ for ILAE1,2 patients were not significantly less than 0.5 (p=0.129, t=-1.15, left tail t-test). Taken together, these group level findings suggest that regions with interictal abnormalities remain after surgery in patients with persistent post-operative seizures. Furthermore, the distinguishability between the resected and spared abnormality (i.e. $D_{RS}$) can discriminate between surgical outcome groups with AUC=0.75.

In contrast, when only using the maximum relative band power in all 62 patients without scoring it against the normative map, patients with persistent seizures (ILAE3+) were not distinguishable from seizure-free patients in any individual frequency band (ILAE1,2) ($\delta$ AUC=0.57 p=0.28, $\theta$ AUC=0.43 p=0.25, $\alpha$ AUC=0.39 p=0.19, $\beta$ AUC=0.52 p=0.78, $\gamma$ AUC=0.45 p=0.70). This result highlights that it is indeed the abnormality relative to the normative map that contains information on epileptogenic tissue, rather than band power *per se*.

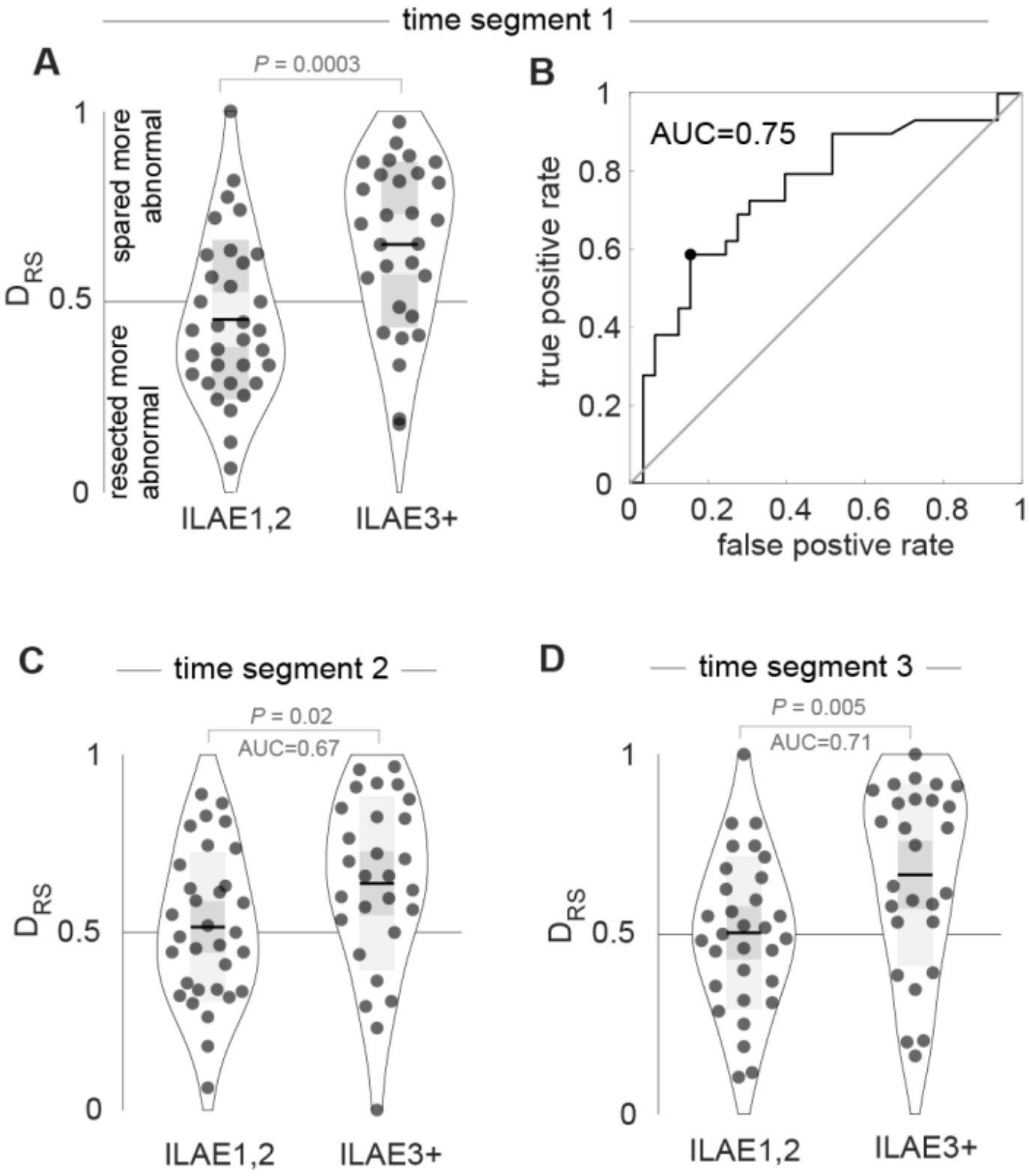

**Figure 4: Interictal band power abnormality distribution in resected *vs.* spared tissue explains post-surgical seizure-freedom. A.** The $D_{RS}$ values, which indicate if resected regions were more abnormal than spared regions, for each patient separated by outcome group. At a group level the resected regions were more abnormal than spared regions in ILAE3+ patients, with substantially and significantly higher $D_{RS}$ values. Each point is an individual patient, black horizontal line indicates the mean, grey box indicates the standard deviation. **B.** Using $D_{RS}$ as a binary classifier with a receiver operator characteristic curve (ROC) allows a calculation of the area under the curve (AUC) = 0.75 to predict ILAE outcome class. **C., D.** Replication of the findings in panel A using other data segments at least 4 hours away from the first data segment.

Finally, for clinical translation, it is also important to assess the robustness of our finding towards the exact interictal segment used. We chose two additional segments of interictal data in the 62 patients, where possible, separated by at least 4 hours and at least 2 hours away from seizures. Repeating the analysis on these two additional segments showed that $D_{RS}$ performed similarly well in discriminating between surgical outcome groups (AUC=0.67, p=0.02 and AUC=0.71, p=0.005, see Fig. 4C,D).

## Robustness of findings

In this section we assess the robustness of our results to various choices of parameters and factors which may influence our abnormality measure $max(|z|)$. We first demonstrate the robustness of our results towards different parcellation schemes, frequency cutoffs, window sizes, segment length, and normative outliers (supplementary analyses 2, 3, 4 and 7). We further demonstrate adequate sampling of the resection by the parcellations (supplementary analysis 9).

Finally, we compared the performance of our abnormality measure compared to interictal spikes as a marker of epileptogenic tissue. Inter-ictal spikes are used as a clinical marker, and are often present for many patients. In figure 5 we investigate if the resection of regions with spikes differentiates outcome groups. The dice similarity used in figure 5 captures the overlap between regions which were resected, and regions with spikes. Unsurprisingly, at a group level, patients had dice values greater than 0.5, indicating that regions with spikes were more often resected. However, the dice similarity overlap does *not* differentiate outcome groups, unlike our abnormality based $D_{RS}$ measure (Fig. 4). Furthermore, if spikes were the main driver of our abnormality metric their presence in the time series would change $max(|z|)$ substantially. However, in supplementary analysis 1 we demonstrate this is not the case. Thus, we conclude that although inter-ictal spikes may be present in patient EEG, their resection does not distinguish outcome groups, and are not the main driver behind our results.

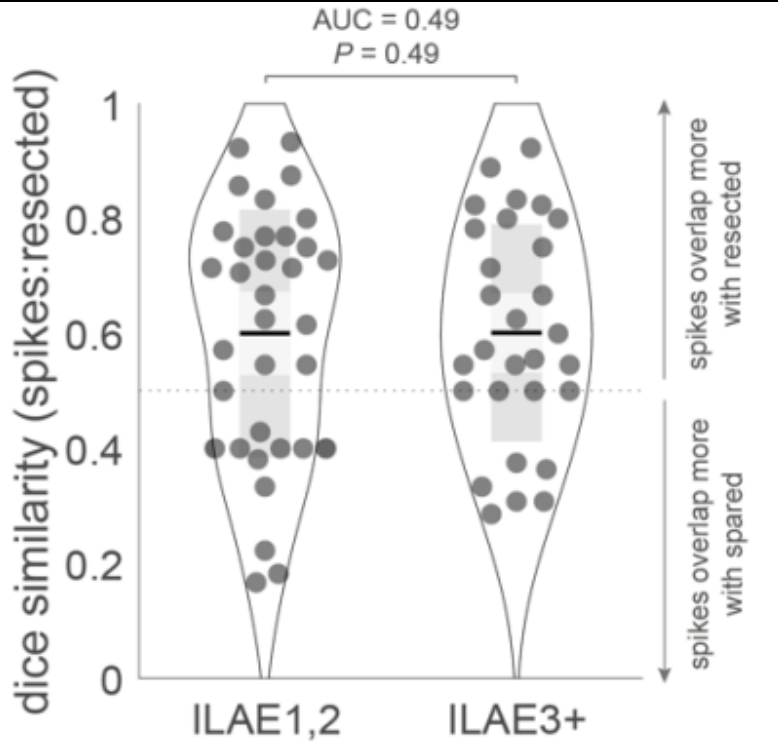

**Figure 5: Resection of interictal spikes does not explain outcome.** Datapoints represent individual patients and indicate the dice overlap of regions containing contacts with inter-ictal spikes and regions which were later resected. Although regions with spikes were more commonly resected (mean dice > 0.5), the effect does not explain outcome.

## Discussion

In this study we derived a normative map of relative band power across the brain using intracranial EEG. The use of normative baselines is commonplace in a wide range of neurology research; however, this approach is rare for invasive modalities such as intracranial EEG. By applying an intracranial EEG normative map in the context of epilepsy presurgical evaluation, we made several key contributions. First, we derived a normative map of interictal band power for different brain regions and frequency bands using the largest dataset to date. Second, we found that we can leverage this normative map to identify regional abnormalities within individual patients. Third, by overlaying abnormal regions with knowledge of resected tissue, we validated our identified abnormalities against surgical outcomes. Finally, we also demonstrated the robustness of our results to the choice of brain parcellation and iEEG segment.

Our normative map, inferred using intracranial EEG, has striking similarities to spectral profiles observed using other modalities such as MEG and scalp EEG.[9,10,12] Some regional frequency-specific neocortical activity patterns are well known, including alpha in parietal regions and beta in motor areas.[30,31] In complement to prior scalp EEG and MEG studies, our analysis also allows the investigation of deep brain subcortical structures with high spatial accuracy. Specifically, we report strong delta power in limbic structures including the hippocampus in agreement with one previous intracranial study.[13] Interestingly, we also found strong delta power in anterior temporal and inferior frontal areas, as reported previously.[11] Given the strong connectivity within limbic, anterior temporal, and inferior frontal areas, including via the uncinate fasciculus, we suggest a potential structural underpinning for the spatial profiles observed in our normative maps. A future comparison of our normative map to a normative white-matter structural connectome could confirm this hypothesis for a given parcellation.

Few studies have used interictal intracranial recordings from multiple subjects to infer a normative brain activity. An early study by[15] investigated data from 15 individuals and mapped spatial profiles of band power estimates. In agreement with our findings, they reported high beta power in motor areas and high theta in superior frontal areas (figure 1), amongst other spatial patterns. Perhaps most similar to our work is the study by,[13] who created a normative map with intracranial data from 106 subjects. The authors suggested that clinical EEGs could be compared to such a map to identify abnormal activity. Our study builds on this literature by creating an atlas from 234 subjects and applying it to an independent sample of 62 patients with epilepsy.

After scoring the epilepsy cohort against the normative map, our goal was to detect abnormalities in interictal EEG activity that may help localise the epileptogenic tissue. To achieve this, we wanted to acknowledge the diversity of possible interictal abnormalities. Therefore, we extracted the maximum absolute abnormality in any frequency band. Our proposed $max(|z|)$ measure is only one of several measures likely to be important for epileptogenic zone localisation, and other dimensionality reduction techniques may be beneficial.[32] Future studies should also investigate the band-specific abnormalities, and relate them to the subject-specific interictal activity patterns (e.g. spikes, slowing, etc.) to aid interpretation. Here, we did not specifically investigate the relationship to particular interictal activity patterns, as we wanted to demonstrate a generalisable framework that can detect interictal abnormalities regardless of the specific nature, pattern, location, or cause of the abnormality. However, it is conceivable that e.g. specific etiologies are associated with specific patterns of interictal abnormality. We also did not control for other factors such as handedness, or eyes open/closed, vigilance state, etc. due to unavailability of this information in our retrospective study design. Future work should investigate the influence of epileptiform activity, as well as various other factors known to impact band power in EEG.

Our study further contributes to a growing literature searching for pre-operative imaging markers of the epileptogenic zone that predict post-surgery patient outcomes (e.g.).[33–40] In general, two main approaches can be used to identify pre-operative markers. The first is to use an entirely data-driven approach. Typically, this strategy involves high dimensional data and feature selection methods.[27,41] However, interpreting the selected features may be challenging. In the present study we instead used a hypothesis-driven approach to identify abnormal regions, which we hypothesised would remain after surgery in patients with persistent seizures. Other studies using hypothesis-driven approaches suggested removing hub regions may explain outcomes,[23,28] whilst using clinical demographics along with imaging has also been suggested.[42,43] Note however, that our proposed measure of band power abnormality may only be a sensitive, but not specific marker of the epileptogenic zone, as band power abnormalities outside of this region may also arise as a functional consequence of the seizures/epilepsy e.g. through propagation of abnormal activity patterns or compensatory mechanisms. Because the exact boundaries of the epileptogenic zone are unknown, even after surgery, it is difficult to determine which abnormal regions definitively fall outside of this region. Notably, though, not all channels with a high abnormality need to be removed to achieve post-operative seizure-freedom (Fig. 3B). In future work, we expect that combining different approaches and biomarkers that are differentially sensitive and specific will yield a translatable and interpretable biomarker of epileptogenic tissue and provide optimal predictions for post-surgical seizure-freedom.[44]

Our normative map approach for localising epileptogenic tissue could complement current clinical analysis of intracranial EEG. Currently, one of the key parts of presurgical evaluation is localising seizure onset. However, resecting the seizure onset zone may not lead to seizure freedom in cases where the seizure onset zone and epileptogenic zone only partially overlap.[6] Furthermore, seizure onset data may not be readily available, the onset location may not be consistent,[45] or the onset pattern may be diffuse for some patients, making it challenging or impossible to conclusively localise the epileptogenic zone using only their seizure data. Thus, to complement this approach, clinicians also evaluate interictal intracranial EEG for abnormalities such as spikes[6] and high frequency oscillations,[46] which may be biomarkers of the epileptogenic zone. As discussed previously, visual inspection of intracranial EEG may miss more subtle frequency changes in neural activity, especially activity that is normal in one region may be abnormal if observed in another. By comparing interictal intranial EEG band power to a normative map, our approach highlights less salient, region-specific aberrations, providing a complementary tool to the traditional visual inspection of ictal and interictal EEG. In our validation, we therefore also opted to compare to surgically resected tissue and subsequent surgical outcome, rather than comparing with e.g. seizure onset zone or irritative zone.

To demonstrate the clinical usefulness of our approach, we showed that the discrimination of surgical outcome groups was robust to the choice of the interictal EEG segment. However, this finding should not be mistaken as evidence that interictal band power abnormality remains stable over time; rather, it simply demonstrates that the predictive power of this measure is not sensitive to abnormality fluctuations on the cohort-level. Nevertheless, there is known variability in interictal dynamics within patients with focal epilepsy. For example, both the rate and spatial patterns of pathological interictal events such as spikes[47–49] and HFOs[50,51] fluctuate during intracranial recordings. Further, interictal band power changes over a range of timescales (see[52] and references therein), and, as a result, band power abnormality also fluctuates over time (see supplementary analysis 10). Future work will investigate the magnitude and timescales of these fluctuations and determine if they hold additional information about epileptogenic tissue. In particular, abnormalities may be more salient following presurgical perturbations such as antiepileptic medication reduction or sleep deprivation,[53] as well as during patient-specific phases of circadian or multiday cycles.[47,48] Our observation that the group-level effect (in predicting surgical outcome) is largely unaltered across different time segment most likely reflects the fact that our sampling in time is random in each patient. In other words, if we can find an "optimal" segment in each patient for detecting their band power abnormalities, we expect our group effect to be even higher. Additionally, like other interictal features,[50,54] temporal changes of abnormalities could also be related to variable seizure features such as seizure onset[55] or evolution[56] within the same patient. Investigating such relationships could reveal additional applications for band power abnormalities, such as predicting seizure features.

Our study has several strengths and limitations. One strength is the sample sizes for both the normative map and epilepsy surgery datasets, which are some of the largest reported in the literature on intracranial EEG. Furthermore, the availability of patient data from other modalities including pre-operative MRI, CT, and post-operative MRI allowed for accurate electrode localisation and delineation of resections. The reproducibility of the normative map across parcellations, and its agreement with existing literature, is also a major strength, providing confidence in our findings. The study's limitations include the retrospective design of the study and the single-site origin of the patient data. Additionally, data regarding the brain state of the patients at the time of recording was not included in the analysis. Future studies could investigate if normative maps and outcome predictions are affected by underlying state changes such as rest, task, or sleep.

Patients undergoing invasive monitoring for surgical evaluation are typically those with the most uncertainty around where to operate, and they subsequently experience poorer outcomes as a difficult-to-treat cohort. Therefore, new ways to use invasive intracranial data are sought after to inform and improve clinical decision making. We envisage, in future, a software tool containing a normative map to which patient data and planned resections are compared.[27] Such a tool would integrate other abnormality metrics from additional modalities including scalp EEG, MEG, or MRI[18] and make predictions of patient outcomes using advanced computational models of brain dynamics.[40,57] Our findings pave the way to the use of normative intracranial baselines for clinical abnormality identification in epilepsy and beyond.

## Acknowlegements

We thank members of the Computational Neurology, Neuroscience & Psychiatry Lab (www.cnnp-lab.com) for discussions on the analysis and manuscript; and Catherine Scott and Roman Rodionov for helping with data organization. B.D. receives support from the NIH National Institute of Neurological Disorders and Stroke U01-NS090407 (Center for SUDEP Research) and Epilepsy Research UK. Y.W. gratefully acknowledges funding from Wellcome Trust (208940/Z/17/Z). P.N.T. is supported by a UKRI Future Leaders Fellowship (MR/T04294X/1). T.O. is supported by the Centre for Doctoral Training in Cloud Computing for Big Data (EP/L015358/1).

# Competing interests

The authors report no competing interests.

# Supplementary Material: Normative brain mapping of interictal intracranial EEG to localise epileptogenic tissue

Peter N Taylor[1,2,*,†], Christoforos A Papasavvas[1], Thomas W Owen[1], Gabrielle M Schroeder[1], Frances E Hutchings[1], Fahmida A Chowdhury[2], Beate Diehl[2], John S Duncan[2], Andrew W McEvoy[2], Anna Miserocchi[2], Jane de Tisi[2], Sjoerd B Vos[2], Matthew C Walker[2], Yujiang Wang[1,2,*,†]

[†]These authors contributed equally to this work.

**Author affiliations:**

1. CNNP Lab (www.cnnp-lab.com), Interdisciplinary Computing and Complex BioSystems Group, School of Computing, Newcastle Helix, Newcastle University, NE4 5TG, UK

2. UCL Queen Square Institute of Neurology & National Hospital for Neurology and Neurosurgery (NHNN), Queen Square, London WC1N 3BG, UK

*Correspondence to: Peter Taylor & Yujiang Wang

Full address: Urban Science Building, Newcastle Helix, Newcastle upon Tyne, UK

Email: peter.taylor@newcastle.ac.uk yujiang.wang@newcastle.ac.uk

**Running title:** Normative iEEG mapping and epilepsy

# Supplementary analysis 1: the impact of interictal spikes

In the following analyses we investigate the impact of interictal spikes on our abnormality measure max |z|. We perform two separate analyses.

Our first spike analysis is as follows. Frequently spiking channels were marked by the clinical team at NHNN (Queen Square) for the UCLH dataset. We investigated (a) if our abnormality measure is different between those regions with regular spikes and those regions without (b), if the resection of regions with regular spikes differed between outcome groups, as our abnormality metric ($D_{RS}$) does, and (c) if the dice similarity (overlap between spikes and resection) is correlated with our $D_{RS}$ abnormality metric (which captures the difference between resected/spared abnormality).

Analysis (a). We first investigated, for the two example patients presented in the main manuscript, if abnormality (max |z|) differs between regions with/without spikes (figure S1A). The rationale being that if our algorithm is driven by spikes, then there will be a clear increase in abnormality in spiking regions. We see in both example patients that such an increase is not present. Indeed, for patient 1216 many of the regions with inter-ictal spikes were actually less abnormal than their non-spiking counterparts. From this analysis we conclude that regions with marked spikes are not necessarily more abnormal than regions without marked spikes in these two patients.

To quantify the difference in abnormality between regions with regular spikes (S) and regions with no spikes (NS) we devised the $D_{S:NS}$ measure. The $D_{S:NS}$ measure is analogous to the $D_{RS}$ metric. Instead of quantifying the difference in abnormality between resected and spared regions it quantifies the difference between the regions with spikes and with no spikes. Difference is calculated as an AUROC as described in methods. A $D_{S:NS}$ value of 1 indicates all regions with spikes are more abnormal than regions without spikes.

The $D_{S:NS}$ values for all patients are presented below (Figure S1B). The values below one for the majority of patients support the statement that interictal spikes are *not* the sole or main driver of our abnormality measure.

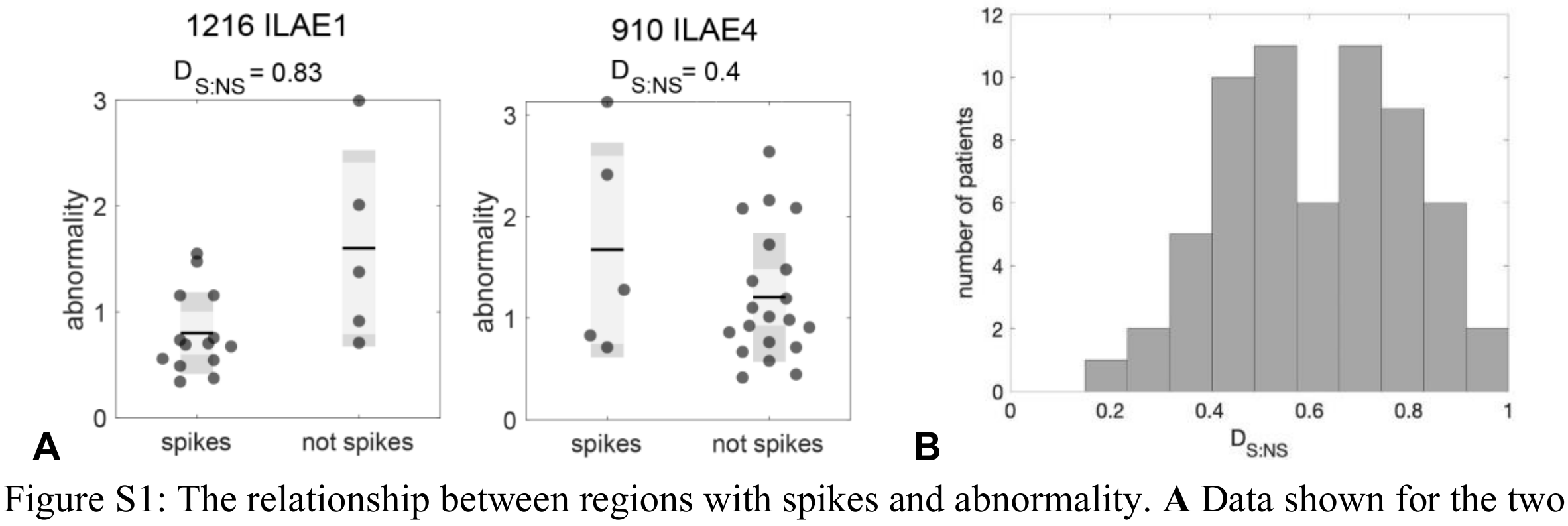

Figure S1: The relationship between regions with spikes and abnormality. **A** Data shown for the two example patients from figure 3 In the main text. Each data point represents an individual region. **B** Histogram of $D_{S:NS}$ values across patients. A value of 1 can be interpreted as all regions with spikes are more abnormal (max |z|) than all regions with no spikes. Almost all patients have values less than 1, with many around 0.5, suggesting the presence of abnormalities (max |z|) present in both regions with spikes, and regions without.

Analysis (b). We used the dice similarity as a measure of the overlap between those regions with spikes, and those regions which were later resected. Where the dice similarity equals one the resected and spiking regions match exactly, where the dice similarity equals zero the spikes overlap exactly with the spared regions. Dice equal to 0.5 represents chance-level overlap. If the removal (sparing) of interictal spikes was sufficient to explain the difference in patient outcomes, then the dice similarity would differ between outcome groups. We show in figure S2 below that dice similarity is not significantly different between outcome groups (p=0.49, auc=0.49). From this analysis we conclude that the resection of spike regions does not discriminate outcomes, unlike $D_{RS}$ based on abnormality (Figure 4, main text). Unsurprisingly, dice similarity values are greater than 0.5 in most patients suggesting that regions with spikes were more commonly removed.

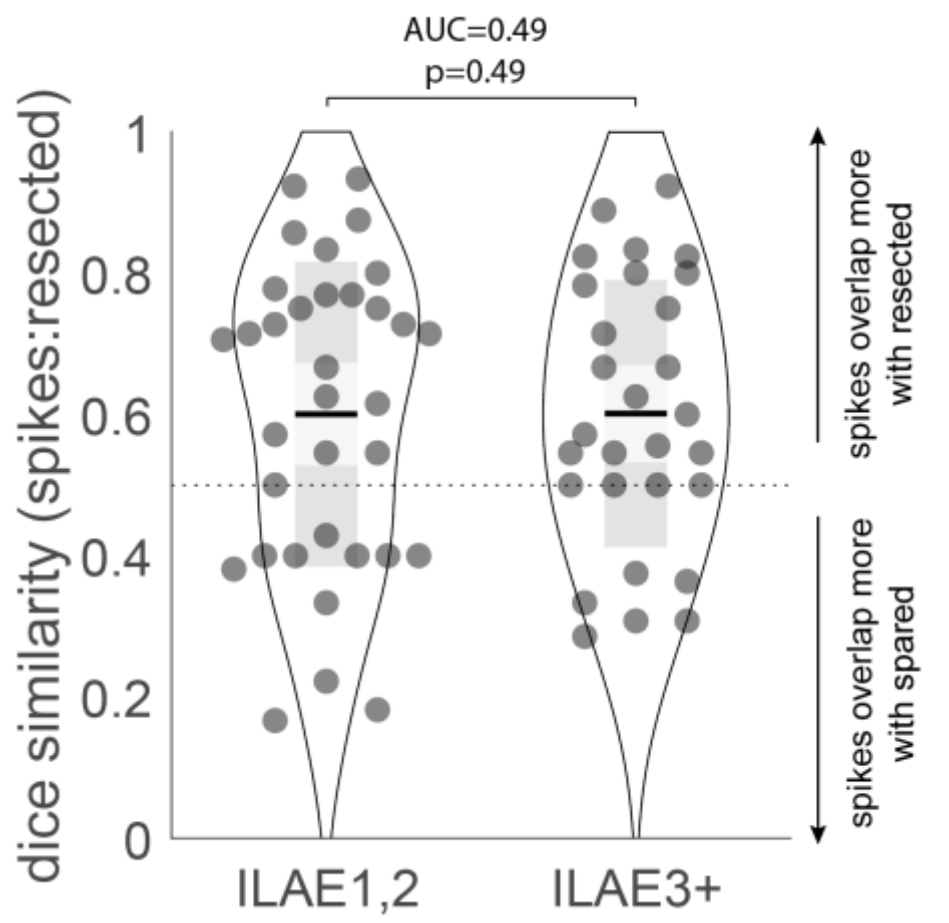

Figure S2: Overlap of regions with spikes and regions which were later resected, quantified using the dice similarity. Each datapoint is an individual patient. High values greater than 0.5 represent patients in which regions with spikes were resected, whereas low values less than 0.5 represent patients in which regions with spikes were spared.

Analysis (c). We investigated if there is a relationship between abnormality-based $D_{RS}$ and spike-based dice similarity. We find no significant association between these two metrics in either outcome group (red and black data points represent individual subjects), or across our cohort as a whole (Figure S3). Thus we conclude that $D_{RS}$, based on abnormality, is not merely reflecting spike overlap with resection.

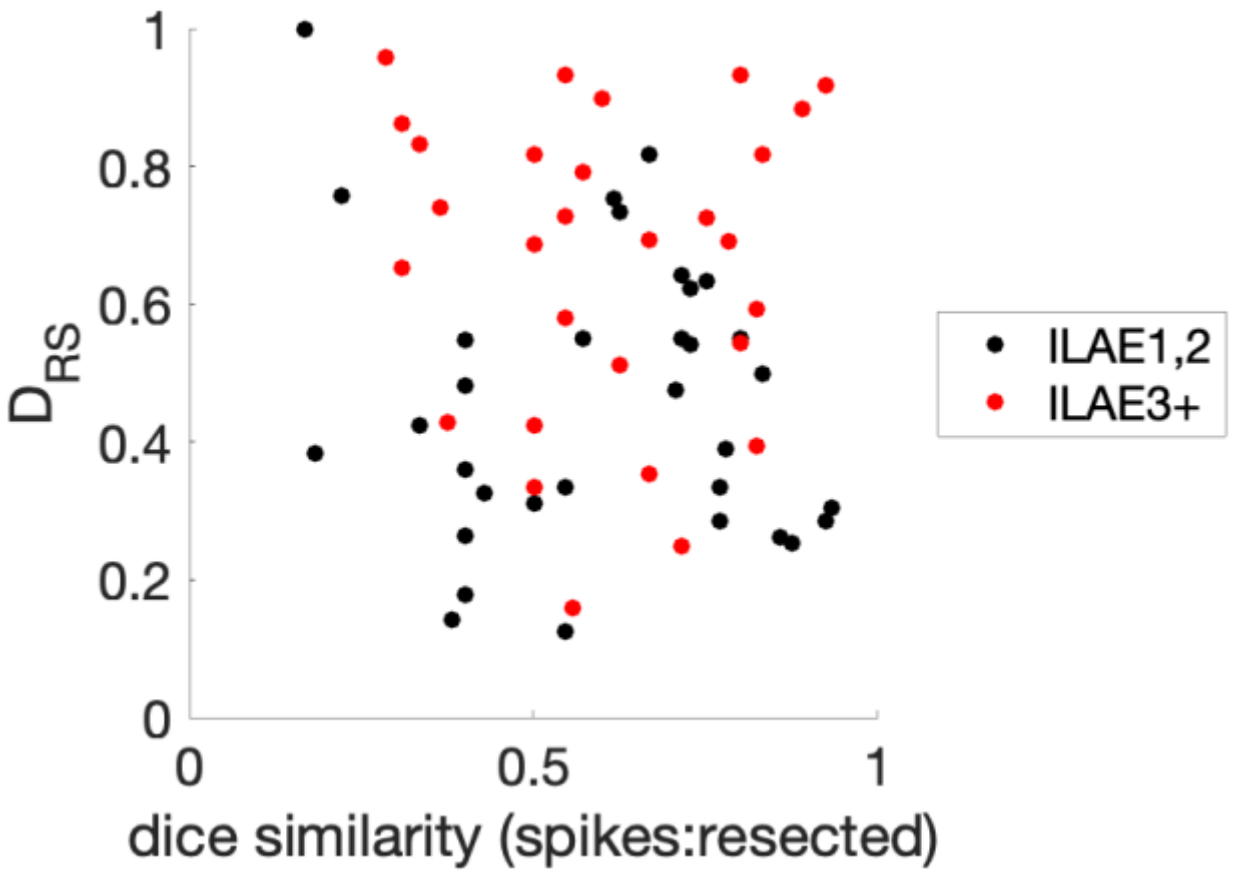

Figure S3: no relationship between $D_{RS}$ (y axis) and dice similarity measure of overlap between regions with interictal spikes and resection (x axis). Each datapoint is an individual patient.

Our second analysis investigates the impact of additional spikes on our abnormality measure (max |z|).

We took the 70 second example time series from an electrode contact placed in the left middle temporal gyrus (part of which is shown in figure 2a in the main text) and artificially added simulated spikes using the approach proposed by Hu *et al*. 2020 (with the spike template derived from Kural *et al*. 2020). After adding simulated spikes we recalculated the relative band power of the modified time series, then computed the abnormality (max |z|). In the process of adding spikes two parameters must be chosen: (i) the number of spikes added (ii) the amplitude of the spike. These parameters are scanned in the plot below. We find that for frequent spikes (9 per minute, 15% of seconds) of very high amplitude (200 mV, equating to 8x standard deviation of the mean of the time series) there is a small change in our abnormality measure (max |z|), an increase of 0.2. This increase is encouraging as we would indeed consider the time-series to be more abnormal, since we have added spikes. Also encouraging however is that the increase is by only 0.2, which does not explain some of the very large abnormalities present in our cohort (see example patients in figure 3 of manuscript with abnormalities >2). We therefore conclude that although the addition of interictal spikes, just like any modification of the time series, will alter the PSD and subsequent max(|z|) the effect is relatively small.

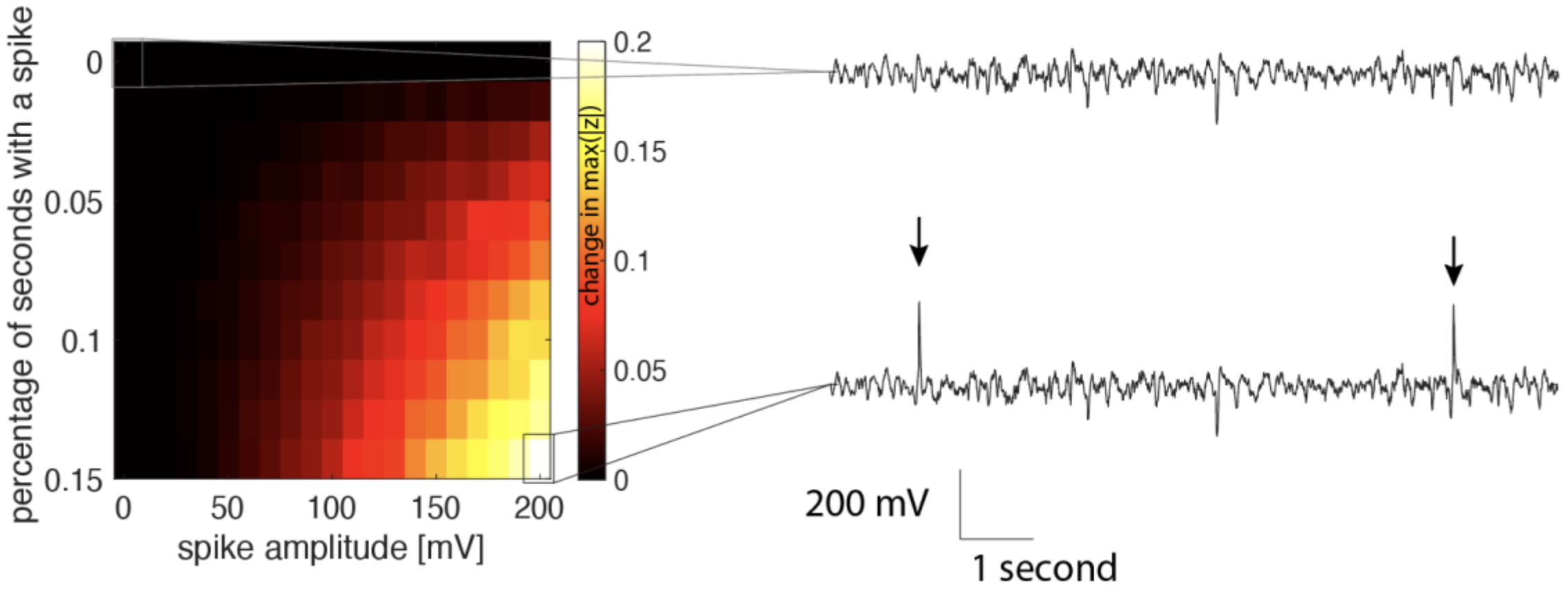

Figure S4: The change in max |z| is plotted for the addition of simulated spikes added to the time series of a region for different amplitudes and frequency (spike rate per minute). For spikes of zero amplitude, or zero frequency, the max |z| is unchanged (black colours in the left plot and top time

series on the right plot). For higher amplitude spikes (200mV) which occur frequently (9 per min, 15%) - lower time series - the max |z| changes by 0.2.

Our findings with the addition of simulated spikes further align with those from the study by Hu et al (2020) who also added simulated spikes to data to investigate their impact. The authors of that study concluded "Simulated spikes in healthy control EEG did not alter network strength or structure."

Summarising these two separate analyses we conclude that although interictal spikes can alter a PSD and subsequent max(|z|) value, such spikes are not the only, or primary source driving our results.

## Supplementary analysis 2: the choice of frequency cut-off

Although many previous EEG studies use typical frequency bands (delta, theta, alpha, beta and gamma), their definition is not always consistent between studies. Here we investigate if our findings are sensitive to subtle changes in the boundary definitions for each band. Table SA1 below shows three examples using different frequency band cut-offs. Row one replicates the original results in the main manuscript. Rows two and three show the findings with slightly different cut-offs. We report that our findings are broadly similar and conclude to *not* be sensitive to subtle changes.

**Table SA1: using slightly different cut-offs for each of the major frequency bands leads to similar results.**

| **Band1** | **Band2** | **Band3** | **Band4** | **Band5** | **AUC** | **P** |
|---|---|---|---|---|---|---|
| 1-4 | 4-8 | 8-13 | 13-30 | 30-80 | 0.75 | 0.0003 |
| 1-3.5 | 3.5-7 | 7-12 | 12-28 | 28-70 | 0.77 | 0.00008 |
| 1-5 | 5-10 | 10-15 | 15-35 | 35-80 | 0.7137 | 0.0014 |

## Supplementary analysis 3: Consistency using different window size & segment duration

In the main manuscript we used a window size of two seconds for Welch's method, over a segment of 70 seconds of data. Here, we investigated the consistency of our results when using different choices for these parameters. Figure S5 shows our findings to be broadly consistent and robust to the choice of window size and segment length. Note that the middle panels in figure S5 are identical and represent the parameter values used in the main manuscript.

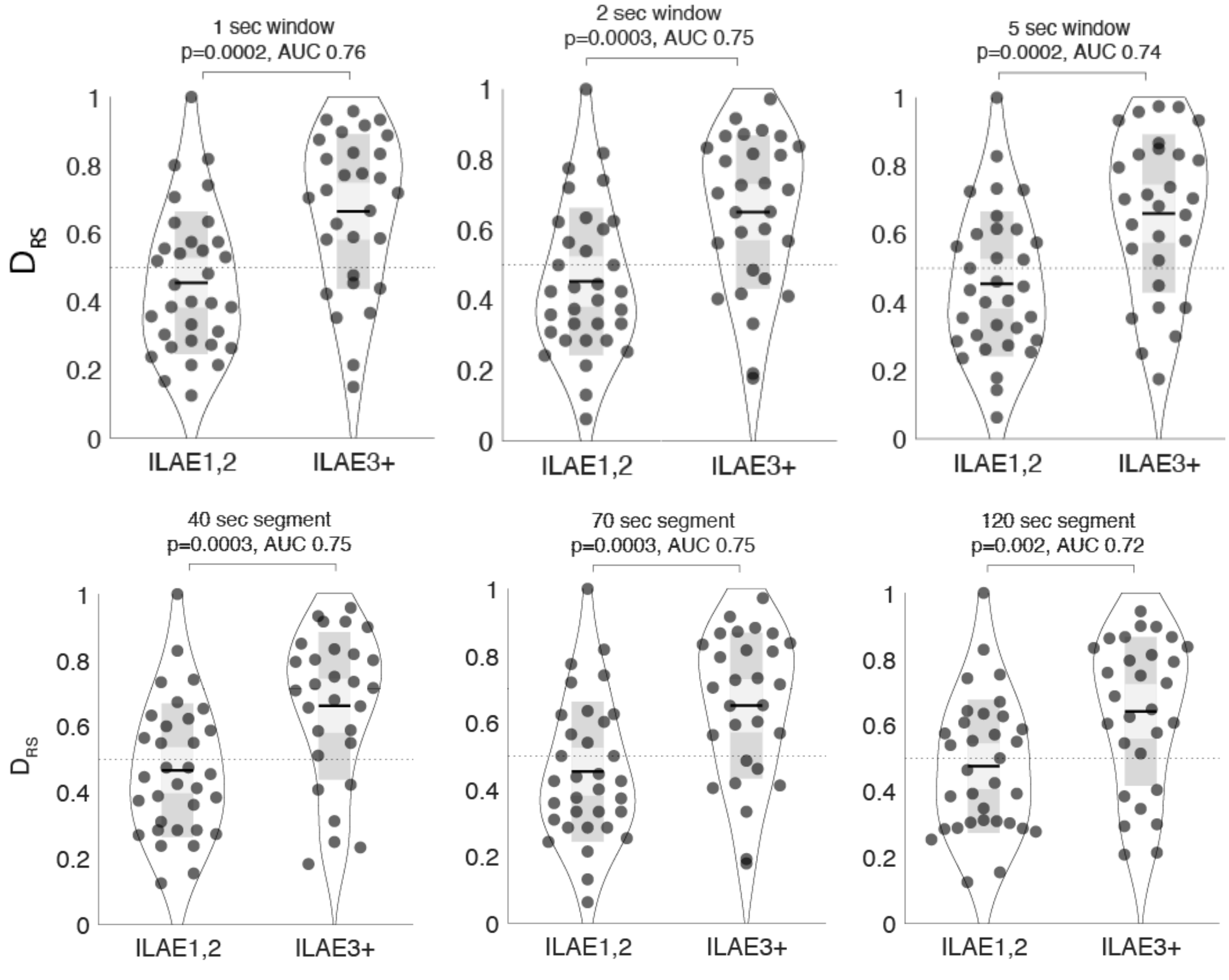

Figure S5: Robustness of results to choice of window or segment length

# Supplementary analysis 4: Consistency across parcellations

In the main manuscript we used a parcellation with 128 regions of interest. Here we demonstrate the consistency of our result across alternative parcellations with different resolutions (figure S6). Both the normative map and the difference between ILAE1,2 and ILAE3+ groups are similar across all parcellations.

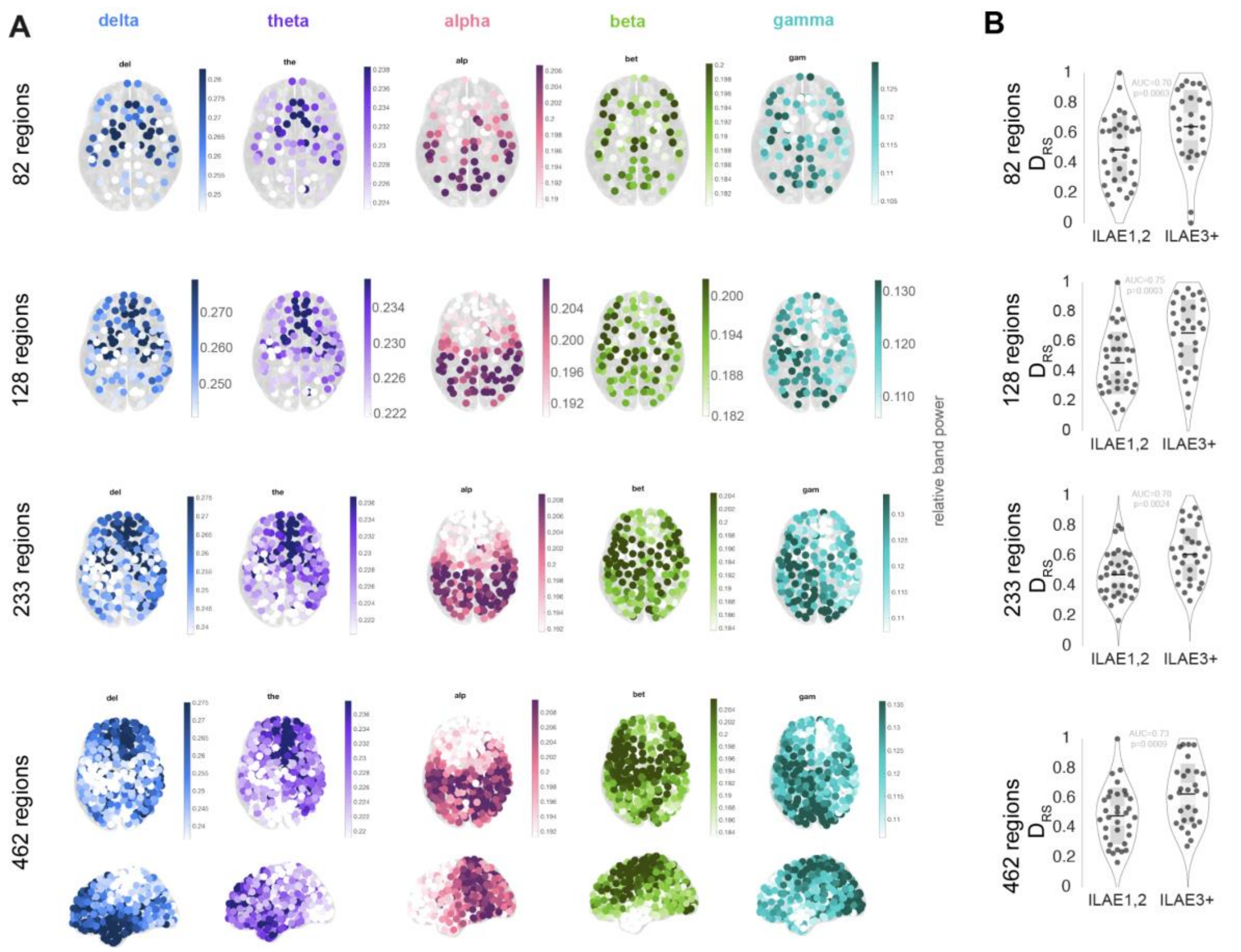

Figure S6: Replication of findings across parcellations **A.** Normative maps replicated for four different parcellations. The second row uses identical data to that shown in figure 1 in the main text. **B.** Replication of surgical outcome findings across parcellations.

# Supplementary analysis 5: Lobe analysis of normative map

To facilitate quantification of the effect shown in figure 1 we present standardised data in figure S7. The standardisation procedure identifies, for each region, the amount by which that region is dominant in a given frequency band above and beyond other regions. Thus, by standardisation we identify which frequency is the strongest contributor to its activity, relative to other regions. Figure S7A shows the identified frequency for all regions, and figure S7B replicates this data for each lobe to assist visualisation. Clear spatial profiles can be observed with the strong parietal alpha activity shown in figure 1 brought through clearly in figure S7B. Through this visualisation other spatial variations in band power are distinctly visible, for example strong beta in frontal motor areas (green), theta in superior frontal areas (purple), and delta in temporal areas (blue) amongst others.

The regional differences in the normative map are visually apparent. To quantify and summarise these spatial variations we show in figure S7C the average relative band power within a lobe for each frequency band. Each lobe has distinctly different profiles and contributors, underscoring the substantial regional variations in band power.

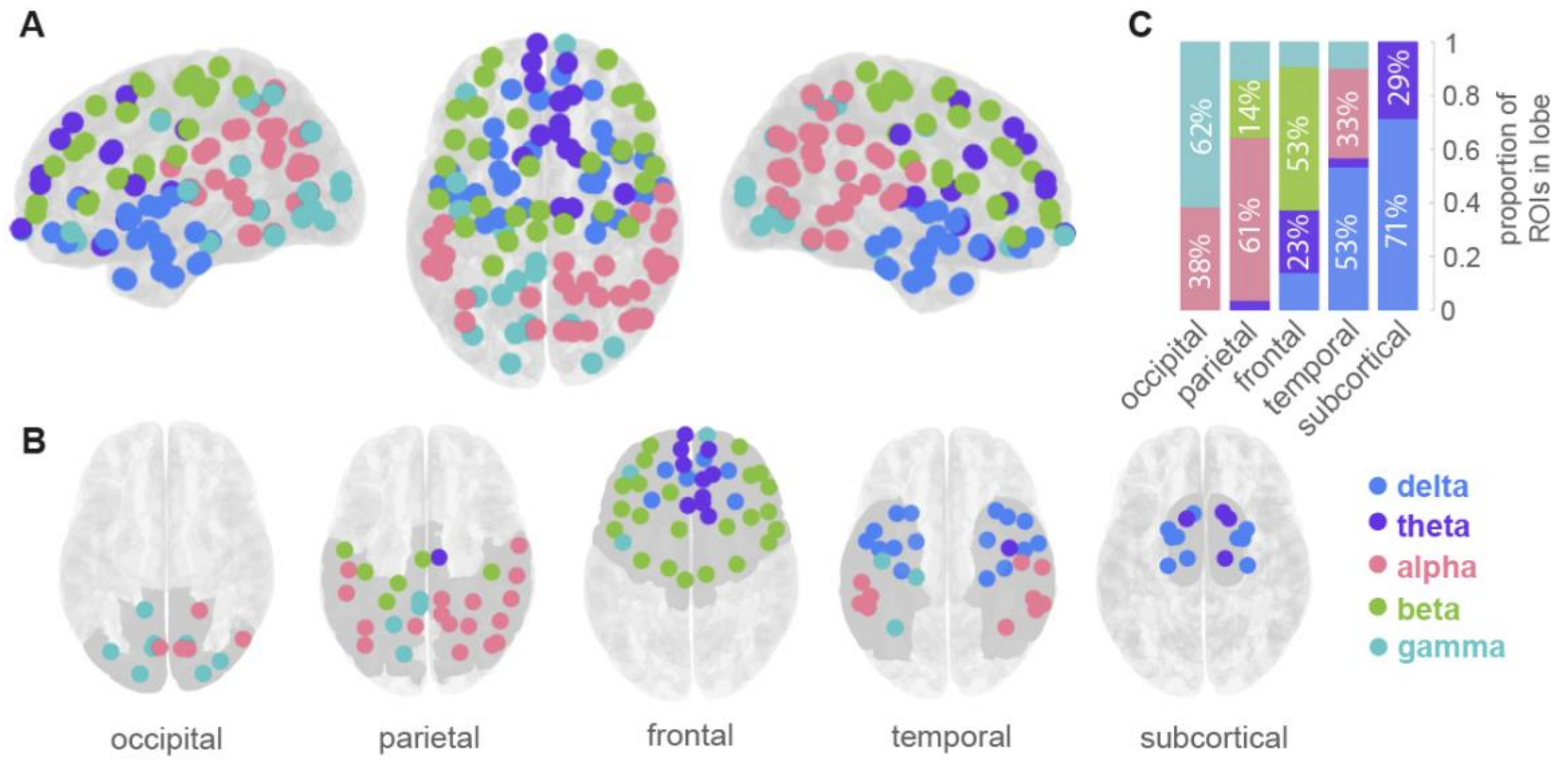

Figure S7: Standardised relative band power, with the maximum frequency band plotted. Standardisation (mean subtract, divide by standard deviation) across regions within each independent frequency band allows all panels in figure 1 of the main text to be placed on the same scale. Panel **A** shows the frequency band which is highest after the standardisation procedure. Note the pink regions in parietal areas which are due to the strong parietal alpha shown in figure 1 of the main text. **B** Replication of panel A showing lobes individually to facilitate visualisation. **C** Proportion of regions within a lobe which are dominant after standardisation for a given frequency. Note that 61% of parietal regions are dominant in alpha activity, thus quantifying the observations in panels A,B and figure 1 in the main text.

# Supplementary analysis 6: spatial distribution of electrode coverage

Implantation of electrodes can differ between all subjects in terms of the location and number of regions covered. In this analysis we show that the coverage is broadly similar between datasets (Figure S8).

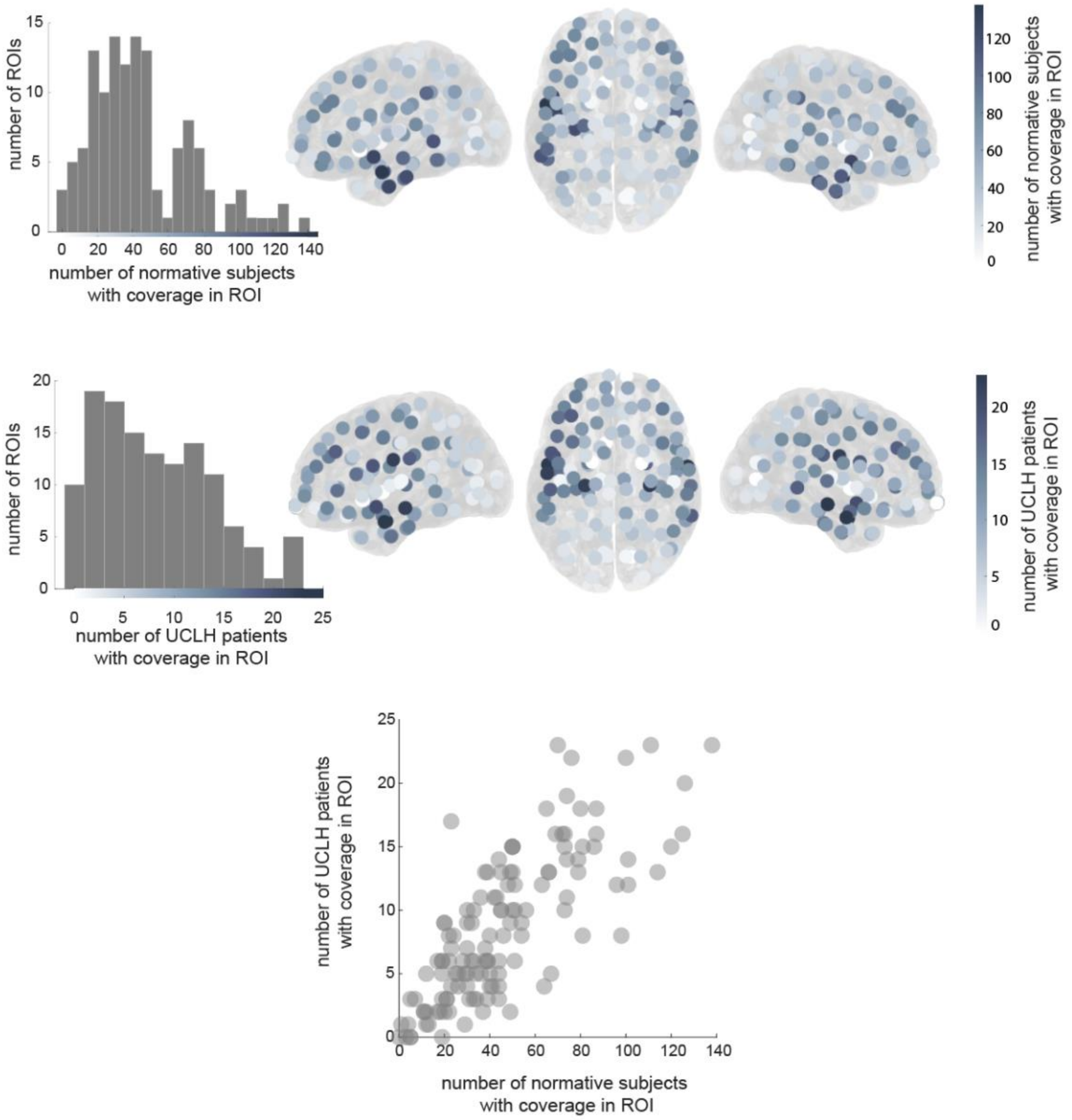

Figure S8: coverage of brain regions in the normative dataset (top panels) and the UCLH patient dataset (middle panels). Lower panel scatter plot shows high similarity in the spatial distribution of coverage between the datasets; each datapoint is a region.

## Supplementary analysis 7: sensitivity to outliers in normative map

In the main manuscript we estimate abnormality in a patient's region as the absolute number of standard deviations (SD) from the mean of the same region in the normative data. However, if the normative data contains outliers they may bias the estimation of the mean in a disproportionate manner. As an alternative in figure S9, we replicate our analysis using the median and median absolute deviation (MAD). We find similar results in both techniques (Pearson rho = 0.92, figure S9).

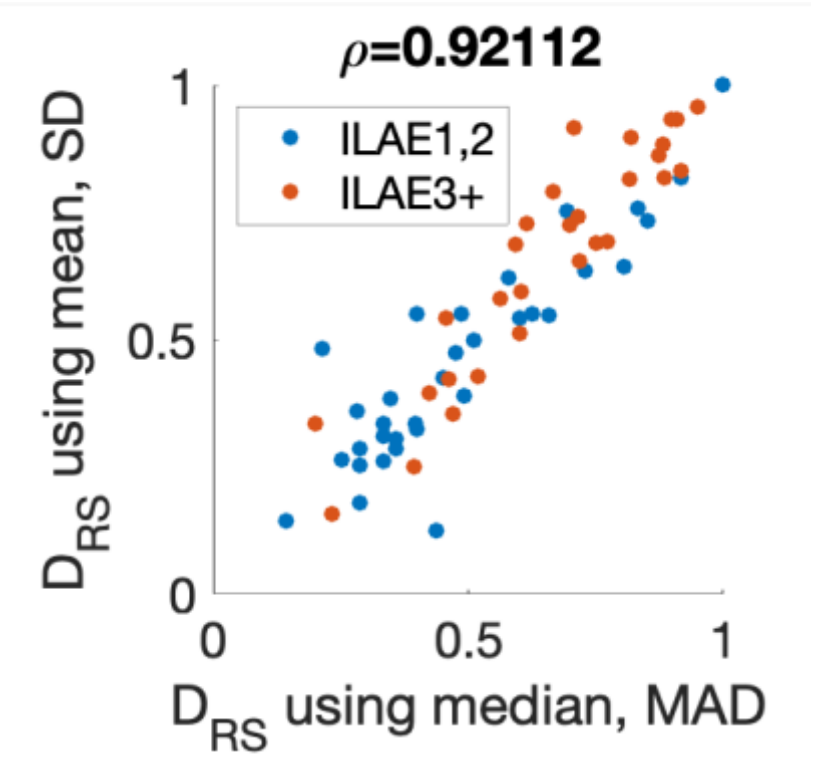

Figure S9: $D_{RS}$ using median & median absolute deviation (MAD) is very similar to $D_{RS}$ using mean & standard deviation (SD).

# Supplementary analysis 8: Individual patient's max |z| abnormality

Here we investigated if particular frequency bands selected as most abnormal (highest max |z|) were more commonly resected regions in patients. In figure S10 we show the frequency bands selected as maximally abnormal for each region as a proportion of all resected regions. As an example, patient 1038 had 7 regions resected. Six of the 7 were most abnormal (had max |z|) in beta (shown as 86% green), whilst one of 7 regions was most abnormal in delta (shown as 14% blue). There is no clear consistency across subjects regardless of outcome group in this individual patient analysis.

Following the previous individual patient analysis in figure S10, figure S11 shows a group analysis which combines all regions for all patients dichotomised by outcome and resected/spared. We observe no substantial difference between any of the four groups represented in figure S11. We thus conclude our results are not driven by one specific frequency band.

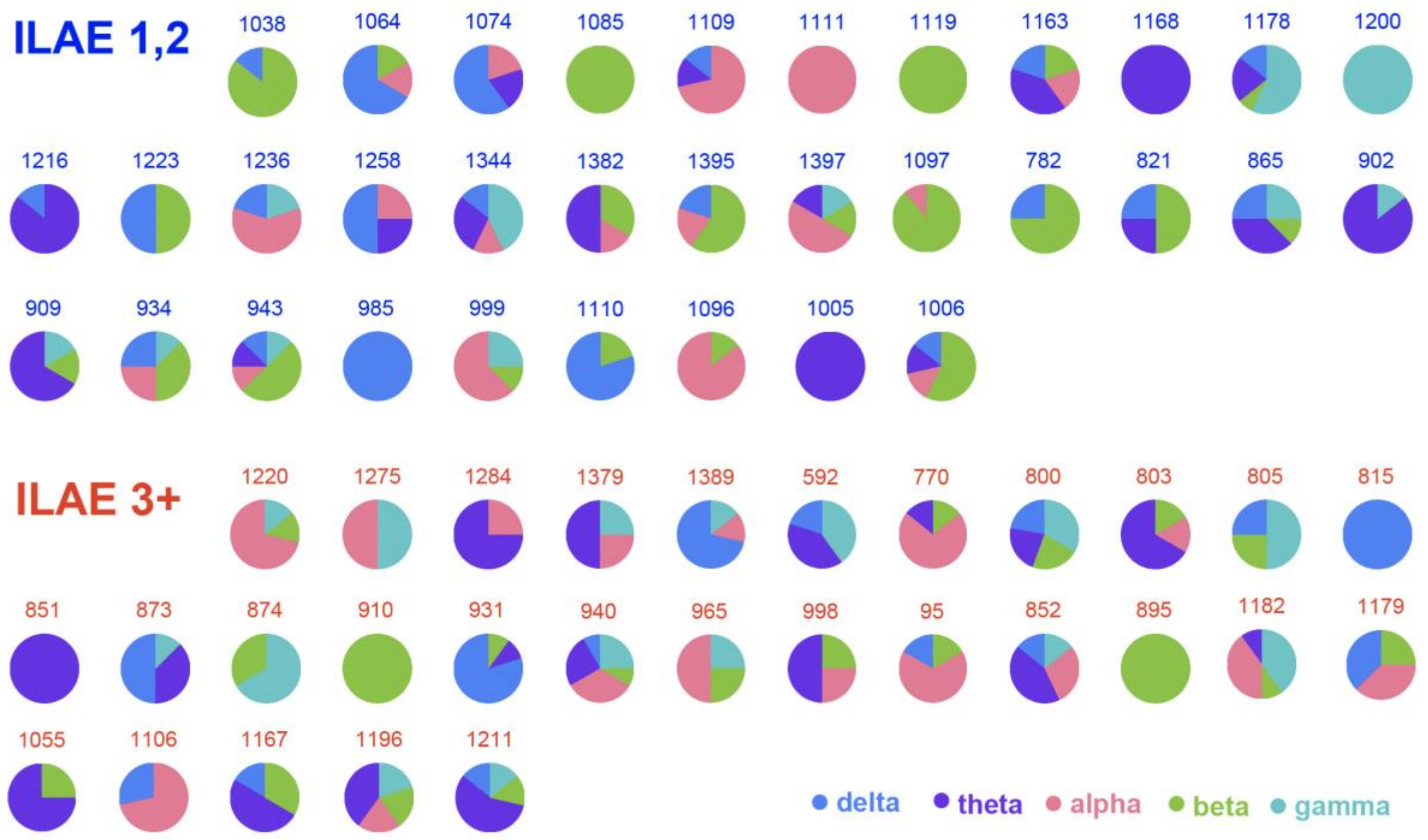

Figure S10: The frequency band found to be most abnormal (have the max |z|) in the resected regions plotted as a pie chart per subject. As an example, patient 1038 had 7 regions resected. Six of the 7 were most abnormal in beta (shown as 86% green), whilst one of 7 was most abnormal in delta (shown as 14% blue). Each chart is an individual patient, the number above the chart is the patient identifier referred to in table S1.

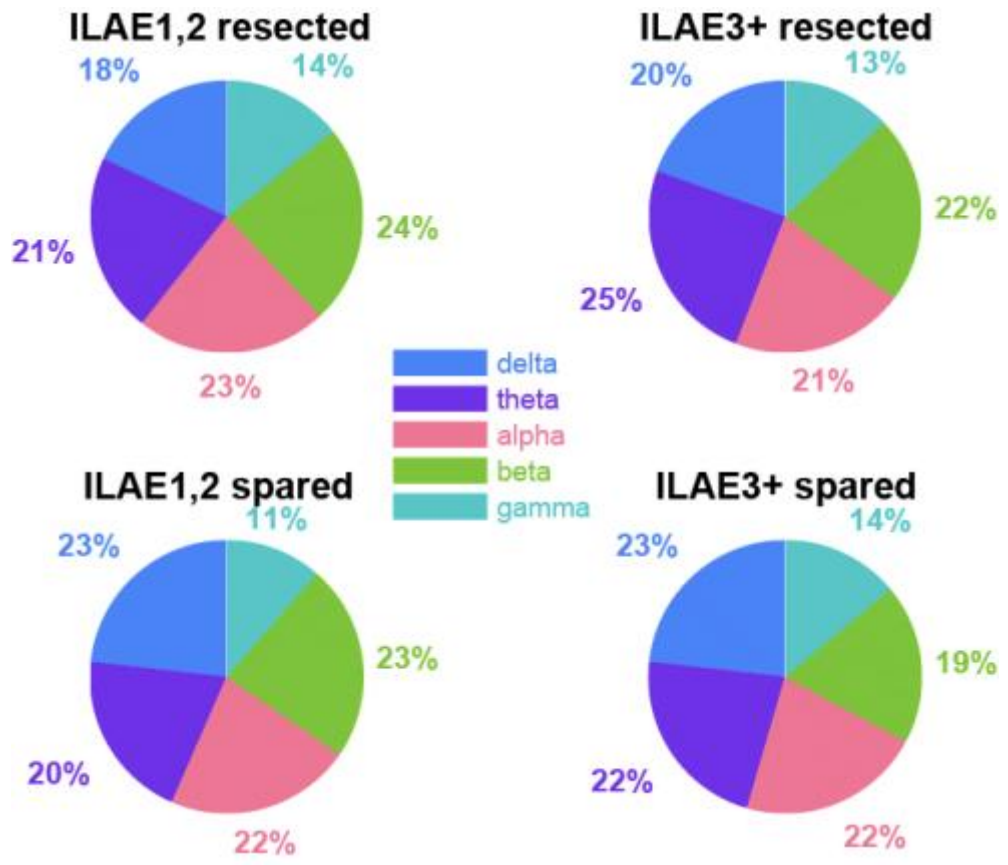

Figure S11: across all subjects the regions' band selected as max |z| is approximately evenly distributed across delta, theta, alpha, beta, and gamma.

## Supplementary analysis 9: Volumetric size of the resection

In this analysis we investigate the size of the resection and the size of the regions of interest for a given parcellation (figure S12). Region volumes (top row) reported below are from the MNI space atlas, and not specific to any individual patient. We find that most regions in all parcellations have smaller volumes than the average resection volume, and hence should produce adequate sampling. However, this will be subject and location-dependent, and in theory parcellation 4 should provide the best sampling of the resected tissue, but will also provide the least confidence in each ROI normative distribution.

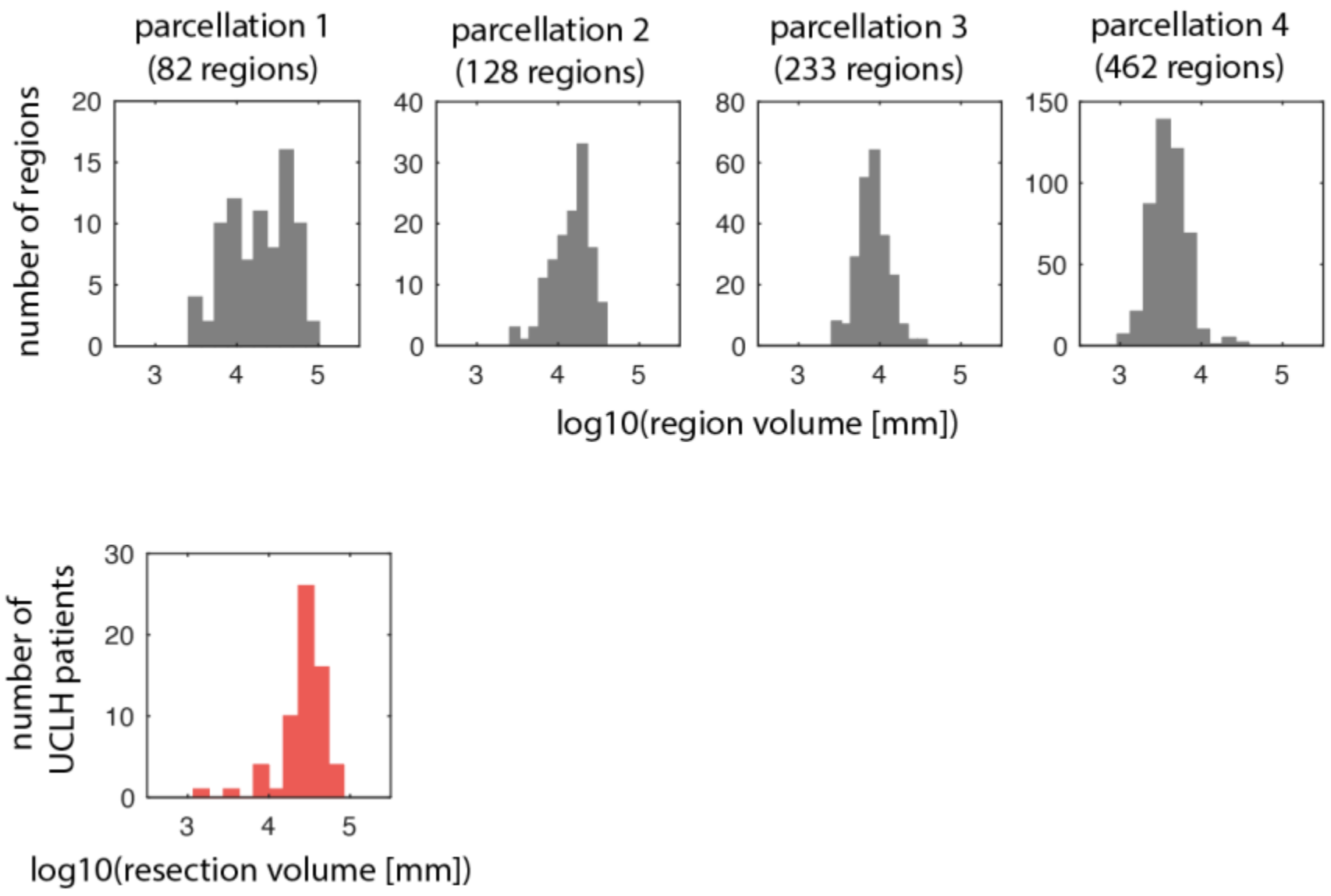

Figure S12: Upper panels, volumetric size of each region in each parcellation. Lower panel, volumetric size of the resection in each patient.

## Supplementary analysis 10: Consistency of $D_{RS}$

In the main text, we presented a consistent AUC regardless of the different time segments we used for each patient. Here we present additional results that clarify the interpretation of the main results. The AUCs presented in the main text measure a group-level effect of $D_{RS}$ being different between surgical outcome groups. However, the consistency of the AUCs across time segments should not to be interpreted as the $D_{RS}$ or the abnormality remaining static over time. Rather, it only indicates that the group-level effect remains regardless of time segment. Below (Fig. S13), we show the actual $D_{RS}$ values for each segment scattered against that of another segment for comparison (each datapoint is a patient). While we can see an overall correlation between the time segments (indicating that the $D_{RS}$ values do not change dramatically from one segment to the next), there is also an amount of temporal variability (as expected - see Discussion).

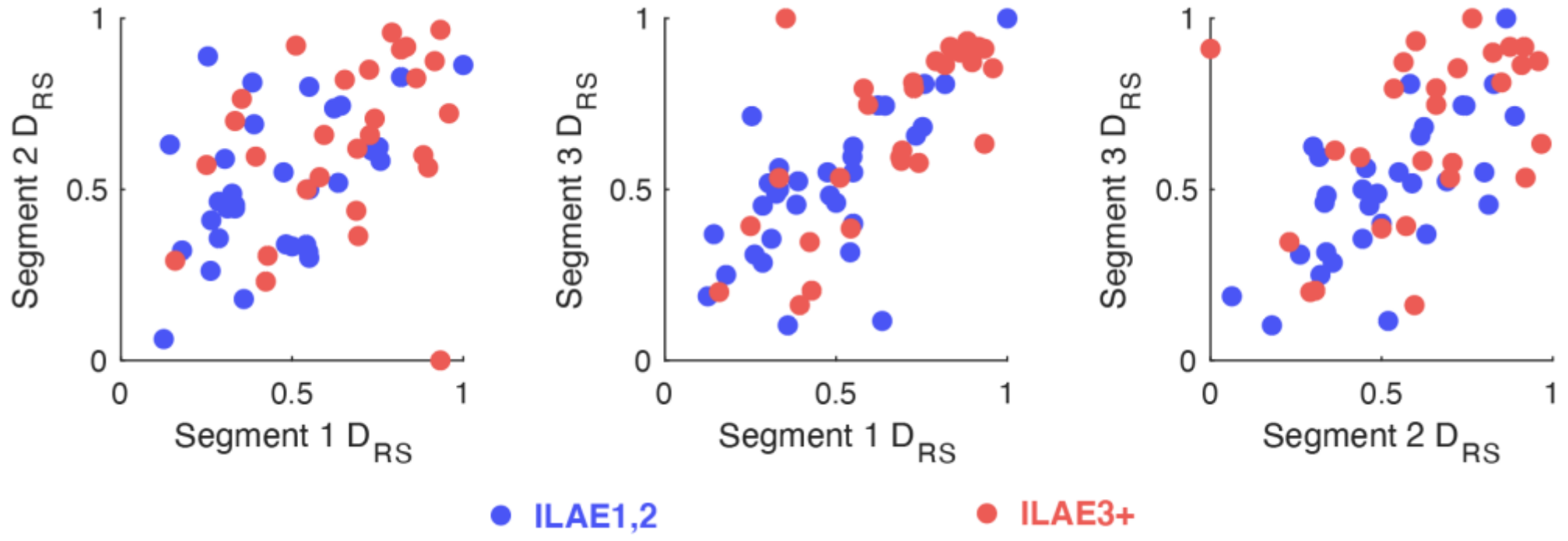

Figure S13: relating $D_{RS}$ values across segments shows some variations, but with a sufficiently large effect size to discriminate outcomes. Each data point is an individual patient.

This result indicates that abnormality levels can fluctuate over time in each patient, and the fact that the AUCs remain similar across time segments reflects the random sampling of different time segments in each patient. In other words, there may be time periods where abnormalities become more salient in each patient, and if we could find and use such time periods *a priori*, then our group-level effect is also expected to increase.